# Unravelling Single Atom Electrocatalytic Activity of Transition Metal Doped Phosphorene

*Akhil S. Nair,[a] Rajeev Ahuja,[b,c] Biswarup Pathak*[a]*

[a]Discipline of Chemistry, Indian Institute of Technology Indore, M. P., India

[b]Condensed Matter Theory Group, Department of Physics and Astronomy, Uppsala University, box 516, SS-75120 Uppsala, Sweden

[c]Applied Matter Theory Group, Department Of Materials and Engineering, Royal Institute of Technology, (KTH) S-10044 Stockholm, Sweden

E-mail: biswarup@iiti.ac.in

**ABSTRACT:** Developing single atom catalysts (SACs) for chemical reactions of vital importance in renewable energy sector has emerged as a need of the hour. In this perspective, transition metal based SACs with monolayer phosphorous (phosphorene) as the supporting material are scrutinized for their electrocatalytic activity towards oxygen reduction reaction (ORR), oxygen evolution reaction (OER) and hydrogen evolution reaction (HER) from first principle calculations. The detailed screening study has confirmed a breaking of scaling relationship between ORR/OER intermediates resulting in varied activity trends across the transition metal series. Group 9 and 10 transition metal based SACs are identified as potential catalyst candidates with platinum single atom offering bifunctional activity for OER and HER with diminished overpotentials. Ambient condition stability analysis of SACs confirmed a different extent of interaction towards oxygen and water compared to pristine phosphorene suggesting room for improving the stability of phosphorene via chemical functionalization.

**TOC GRAPHICS**

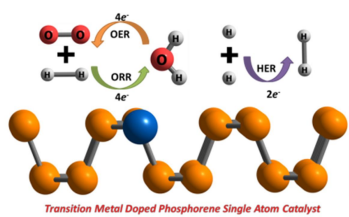

Endeavors to curb the catalyst loading without dwindling the activity has been a far-reaching goal in the area of heterogeneous catalysis.[1-3] Attempts towards realization of this goal along with the development of nanotechnology have explored catalysts of different dimensions such as nanoclusters, nanowires, 2D sheets and so on. The very recent decades have witnessed the emergence of single atom catalysts (SACs), where the atomically dispersed metals catalyzing the reactions and thereby turned out to be the quintessential examples for the green chemistry concept of 'atom economy'.[4,5] A vast number of categories of SACs have been explored so far by varying the supports such as metal oxides, two-dimensional nanosheets, clusters, metal organic frameworks and so forth.[6-9] Apart from the stabilization effects of single atoms anchored on the supports, SACs feature for their compelling electronic, structural and chemical properties emanated out of the synergistic interaction between the support and the metal atom offering extraordinary catalytic properties. For example, Zhang and coworkers demonstrated that Pt atoms uniformly dispersed over iron oxide ($Pt_1/FeO_x$) support exhibited a CO oxidation activity higher than the nanoparticle counterpart.[10] This revolutionary study has been followed by the discovery of a number of $M/NO_x$ (M=Ir, Pt, Au; N=Ti, Fe, Ce) SACs identified as potential catalysts for water-gas shift reaction.[11-13] In order to address the expeditious energy crisis confronted by the modern world, development of efficient electrocatalysts for reactions of energy storage such as oxygen reduction reaction (ORR), $CO_2$ reduction reaction (COR), oxygen evolution reaction (OER) and hydrogen evolution reaction (HER) is an area of utmost research interest. Since the primary catalysts used at an industrial level for these reactions are noble expensive metals such as Pt, Pd, Ir, Rh etc., the evolution of SACs has set off enormous attempts to reduce the catalyst loading by retaining/improving the activity of one-atom based catalysts. So far, the combined experimental and theoretical studies have unraveled a number of transition metal based SACs with pristine and vacant graphene type materials, nanoparticles and so forth.[14-17] A continuous research is being carried out to disclose novel SACs with optimal metal-support combination for excellent activity.

Few layer bulk phosphorus, particularly monolayer phosphorene has been proposed as a material of immense interest owing to the exquisite properties such as unique anisotropic electronic and transport properties, high mobility, turn off ratio and tunable direct band gap.[18-20] These properties have enabled phosphorene to be employed for diverse set of applications such as gas sensing, transport devices, optoelectronics and so on.[21-23] The major obstacle in the practical utilization of phosphorene is its instability at ambient conditions, the origin of which has been attempted to be understood at atomic scale.[24,25] However, the recent studies have been able to synthesize air-stable few layer black phosphorous (BP) by layer by layer thinning of bulk black phosphorous and by pulsed laser ex-foliation.[26,27] Such air-stable phosphorene nanosheets have been also identified with preferred catalytic activity towards electrochemical reactions. In a pioneering work by Zhang and co-workers, few layer BP nanosheets have been identified to be potential catalysts for OER with retaining the structural stability over 1000 potential cycles.[28] The linear relation between the thickness of BP nanosheets and the electrocatalytic activity suggested by the authors recommends the plausibility of monolayer phosphorene to perform as an efficient electrocatalyst. However, Xue et al. through a theoretical study have shown that enhanced catalytic activity for ORR or OER can be derived from phosphorene only with inducing modifications such as vacancies and hetero-atom doping.[29] Therefore, the selective functionalization of phosphorene draws considerable attention as it can deliver significant electrocatalytic activity by tuning the interaction with reactive intermediates as well as enhancing the stability by modifying the electronic structure.

Motivated from the aforementioned ideas, we attempt to investigate the activity of transition metal doped phosphorene SACs (TM-Ph) towards major electrocatalytic reactions such as ORR, OER and HER based on first principles calculations. The final transition metals from 3d (Fe, Co, Ni, Cu), 4d (Ru, Rh, Pd, Ag) and 5d (Os, Ir, Pt, Au) series are considered owing to their catalytic applications for various processes at the bulk level. The structural, electronic and thermodynamic aspects of the catalytic activity are systematically studied and an activity screening among the considered catalyst species is carried out. The origin behind the activity trends observed is scrutinized to derive suitable activity descriptors and related parameters. The active candidates are further analyzed for their stability under ambient conditions by an atomistic approach. The results are expected to provide useful thumbnails for the futuristic efforts in enabling phosphorene as an efficient support for SACs.

The density functional theory (DFT) calculations are carried out using VASP (Vienna Ab Initio Software Package)[30] Gradient corrected approximation and Perdew-Burke-Ernzerhof (GGA-PBE) exchange-correlation functional is employed with projected augmented wave method (PAW) for describing the periodic boundary conditions.[31,32] The study by Hashmi and Chan has reported the exactness between the magnetic ground state prediction between DFT-GGA and DFT-U for the 3d TM-Ph systems.[33] Hence we have avoided considering Hubbard calculation in our study. A kinetic-energy cutoff of 470 eV with K-POINT grid of 5x5x1 for brilliouin zone sampling are used for the energy calculations and 15x15x1 is used for calculating electronic structure calculations. The dispersion interactions are accounted by using DFT-D3 method.[34] All the calculations are spin polarized and without any symmetry constraints. An implicit solvation model implemented in VaspSol is used to simulate the electrolyte conditions.[35] Crystal orbital Hamilton population (COHP) analysis has been carried out using Lobster programm.[36] A 4x4x1 supercell with 48 atoms is used to model the phosphorene with the transition metals substitutionally doped by replacing one phosphorous atom at the bridge site constituting an atomic doping percentage of 2.708% (Figure 1). The doped TM atoms are observed to be possessing variable geometric features across the series which are given in Table S1. While the bond lengths of TMs with P atoms (TM-P1/P4) along the zigzag direction remain close to the pristine values, (2.17-2.47 Å), the average TM-P bond length along armchair directions (TM-P2) differ considerably across the TM series (2.5-3.6 Å) with a closeness observed for same group metals. The incorporation of TMs is found to be associated with a diminished buckling as compared to pristine phosphorene which is determined from the interplanar distance (P1-P3) between the outer and basal planes. Out of all the TM-Phs considered, only Fe and Ni doped systems are magnetic with a total magnetic moment of 1 μB.

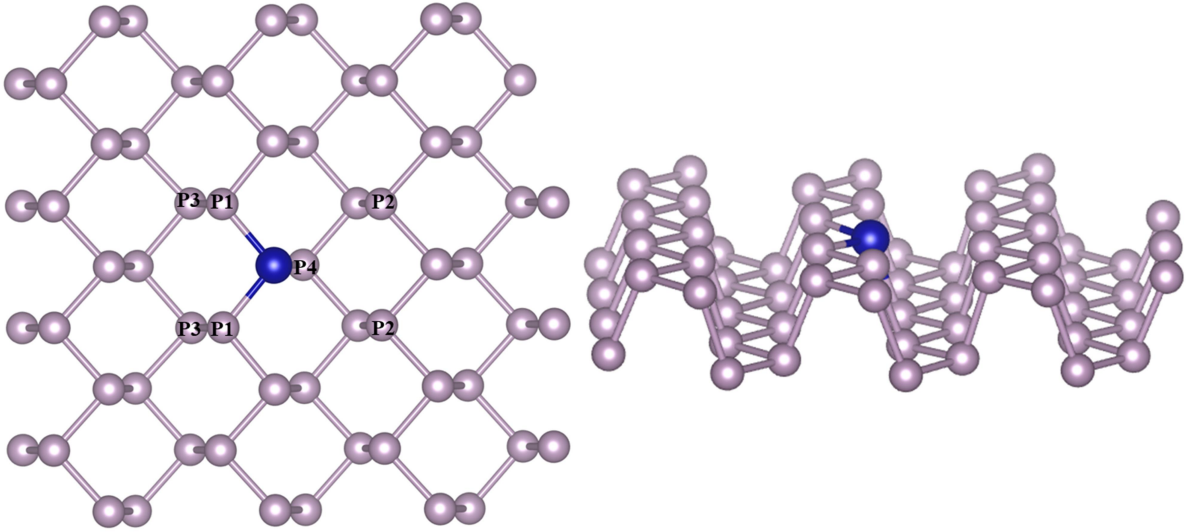

**Figure 1.** Transition metal doped phosphorene model (a) top view and (b) side view. P atoms differing in coordination with respect to TM atom are labeled. Blue and violet colors represent TM and P atoms, respectively

The energetics of phosphorene mono-vacancy formation, the perquisite factor for TM doping is analyzed by calculating the formation energy by,

$$F_E = E_{Ph-Vac} - E_{Ph} - E_P \qquad (1)$$

where $E_{Ph-Vac}$ is the energy of vacancy induced phosphorene, $E_{Ph}$ is the energy of pristine phosphorene monolayer and $E_P$ is the chemical potential of phosphorous calculated from bulk phosphorous. The calculated formation energy of monovacancy in phosphorene is 2.094 eV which is very less compared to the single vacancy formation energy reported for grpahene.[37] Therefore the vacancy inducing can be expected to be experimentally achieved via advanced experimental methods like chemical exfoliation technique. This is followed by an analysis of TM binding energy which is calculated as,

$$E_{BE} = E_{TM-Ph-Vac} - E_{Ph-Vac} - E_{TM} \qquad (2)$$

where the terms from the left to right indicates the energy of TM-doped vacancy created phosphorene, vacant phosphorene and the chemical potential of TM atom determined from its bulk counterpart, respectively (Figure 2a).

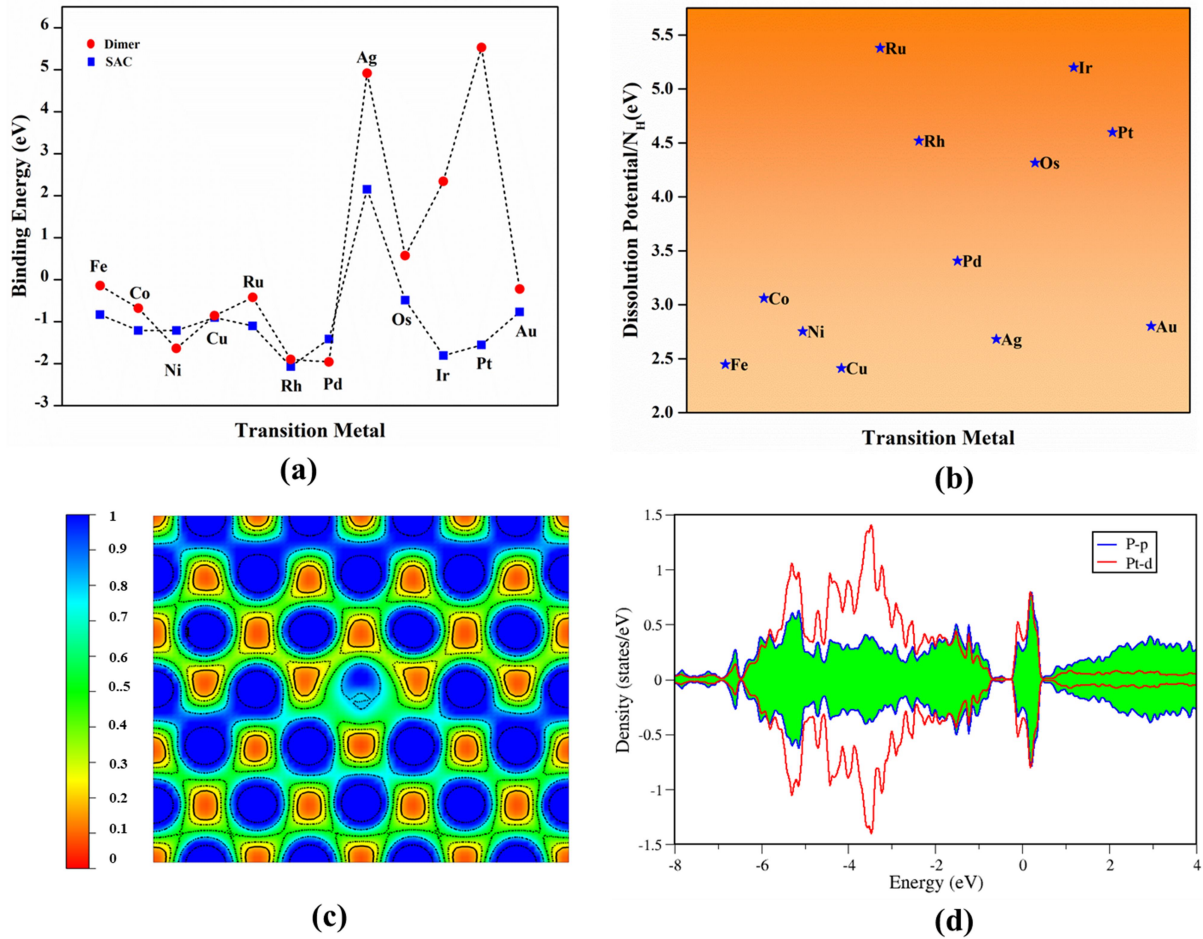

**Figure 2.** (a) Binding energy trends of transition metal single atoms and dimers, (b) Dissolution potential of SACs per number of protons ($N_H$), (c) Electron localization function and (d) Projected Density of States of prototype Pt-Ph system.

All the TM-Ph based systems except Ag-Ph have been found to bind strongly and hence possess the required thermodynamic stability to be useful as catalyst materials. The structural integrity and tolerance towards electrochemical conditions are crucial parameters in determining the long-term activity of electrocatalysts and hence to be addressed in detail. Within this consideration, the relative binding energy of TM-dimer on phosphorene is determined to analyze the plausibility of aggregation of doped TMs in TM-rich condition (Figure 2a: red panel) and compared with single-atom binding scenario. All the TM-dimers are found to be weakly binding compared to single TM atom except for Ni, Pd undergoing dissociation, suggesting a reduced tendency towards aggregation for the SACs. Furthermore, the sustainability of TM-Ph SACs under electrochemical conditions is scrutinized by dissolution potential of TMs calculated by replacing TM atoms with protons followed by the work of Singh et al.[38] as the free energy of the reaction;

$$\text{TM-Ph} + n\text{H}^+ \rightarrow n\text{H-Ph} + \text{TM}^{n+} \tag{3}$$

with nH-Ph indicating TM vacancy occupied by n hydrogen atoms according to the standard oxidation state of the metal. Observed positive dissolution potentials (Figure 2b) per number of protons subjected to the oxidation state of the metals for all the TM-Ph systems convey that they are not prone to dissolution during subsequent protonations. A topological analysis is carried out based on electron localization function (ELF) (Figure 2c) to understand the nature of bonding between the TM and surrounding P atoms. The localization of electron pairs at the atomic centers is revealed by the maximum ELF value close to 0.40 infers a dominancy of ionic bonding character. The projected electronic density of states of all the TM-Ph systems are inspected and that of Pt-Ph candidate for which a high binding energy and dissolution potentials are observed, is shown in Figure 2d as a prototype. The strong hybridization observed between the d-orbitals of TMs and the p-orbitals of coordinated P atoms near to the Fermi level explains the origin of electronic stability of TM-Ph systems. Hence the stability of TM-Ph based SACs is confirmed and hence are further screened for their catalytic activity.

Owing to the similarity between the reaction intermediates involved, we attempt to first study the ORR characteristics of TM-Ph SACs and utilize the useful thumbnails to describe the OER activity trends. The adsorption configuration of ORR intermediates is given in Figure 3(a-e). ORR is initiated by $O_2$ adsorption, the details of which are given in Table S2. The absence of local minima with $O_2$ physisorbed on the TM atoms ensures favourable binding to the single atom center rather than P atoms where physisorbed $O_2$ configuration is observed which is discussed in the later section. A characteristic linear adsorption mode is observed for all the TM-Ph SACs with an average vibrational frequency of 1100-1300 $cm^{-1}$ except Pd-Ph adsorbing in a di-sigma mode. The $O_2$ binding is highly localized at the metal centers in TM-Phs with a lesser $O_2$ binding strength compared to the P atoms of pristine phosphorene among which the highest interaction is evidenced for Fe-Ph with a binding energy of -1.77 eV. Corresponding partial density of states (PDOS) analysis (Figure S1) of Fe-Ph shows strong hybridization between p states of $O_2$ and d-states of iron retaining with a magnetic moment of 3 µB indicating an additive coupling between Fe and triplet $O_2$ magnetic moments. Top site O* adsorption occurs on the TM centers except the end series metals such as Ni, Cu, Pd, Ag and Au where a bridge type adsorption with P2 atoms occurs. OH*, OH* and H* adsorptions also occur in the top mode across all the SACs which follows their typical adsorption configurations over the (111) facets of transition metal surfaces. The local adsorption induced structural changes quantified in terms of deviation of TM-P bond lengths in Figure S2 confirms structural deviations within 0.1 Å change except for Cu and Au based SACs for which a relatively large bond length variations are present. For Ni and Pd, a strong H* binding causes the TM to translate towards the lower layer of TM-Ph and hence leads to distortion of the system. Therefore these (Ni and Pd) SACs are not considered for studying the HER activity.

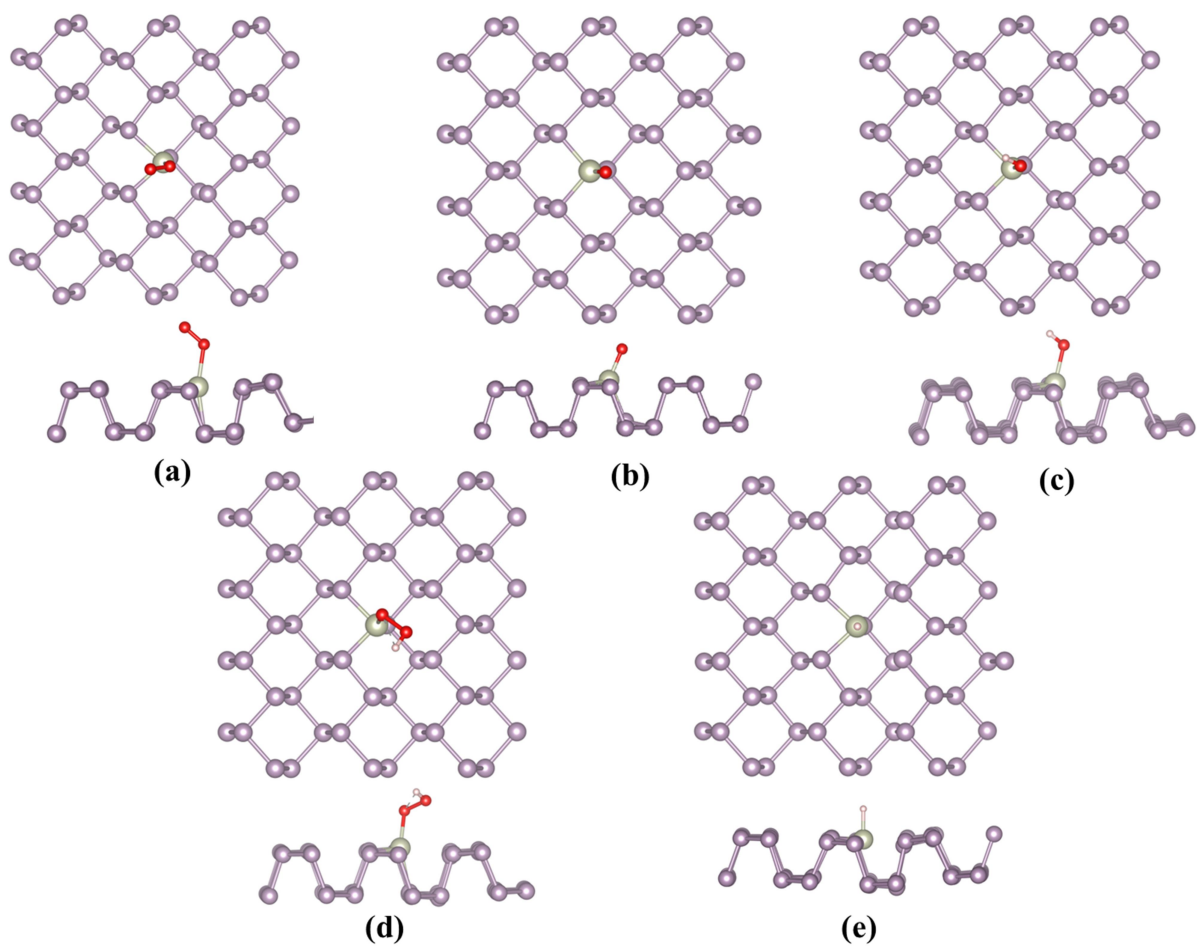

**Figure 3.** Most stable adsorption configurations of ORR/OER/HER intermediates, (a) $O_2$*, (b) O*, (c) OH*, (d) OOH*, (e) H*

Since ORR/OER is a multistep reaction with intermediates O*, OH* and OOH* involved, it is preferable to search for an activity descriptor especially based on one intermediate so that parametric space for catalytic activity can be simplified in number of degrees of freedom and a general trend can be expected. Having this as objective, we have investigated the scaling relations between the adsorption free energies of the intermediates which are attributed to be the fundamental origin for the diminished experimental onset potential for ORR (Figure 4). The adsorption free energies are calculated with reference to the water from the following equations;

$$\Delta G_{O*} = G_{O*} - G_* - [G_{H_2O} - G_{H_2}] \qquad (4)$$

$$\Delta G_{OH*} = G_{OH*} - G_* - [G_{H_2O} - 1/2 G_{H_2}] \qquad (5)$$

$$\Delta G_{OOH*} = G_{OOH*} - G_* - [2G_{H_2O} - 3/2G_{H_2}] \qquad (6)$$

with zero-point energies and entropies calculated from harmonic oscillator approximation along with the DFT calculated binding energy. The interdependency between intermediate adsorption energies is analyzed via scaling relations is shown in Figure 4 which have been identified to be as pivotal factors in determining the electrocatalytic activity.[14-17]

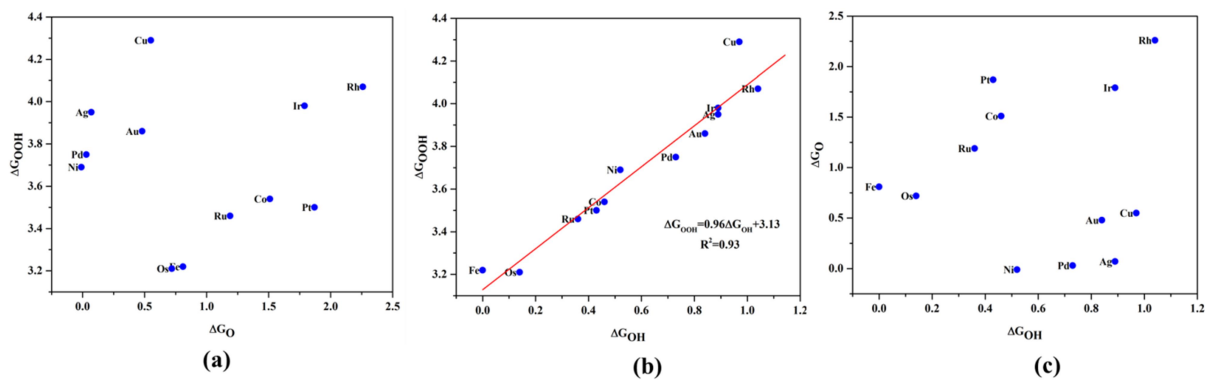

**Figure 4.** Scaling relation of ORR/OER intermediates, (a) $\Delta G_{O*}$ vs $\Delta G_{OH*}$, (b) $\Delta G_{OOH*}$ vs $\Delta G_{OH*}$, (c) $\Delta G_{O*}$ vs $\Delta G_{OOH*}$

A surprising observation is the consistency in linear scaling of $\Delta G_{OOH} = 0.96 \Delta G_{OH} + 3.13$ with a large determination coefficient between OH* and OOH* as seen for transition metal surfaces, whereas almost negligible scaling found for OH* vs O* and OOH* vs O*. This suggests a breaking of scaling relationships across the TM-Ph series which is a sufficient but not an inevitable reason for the catalytic trends. The involvement of nearby phosphorous atoms in bonding with the O atoms can be expected to be providing a difference in binding of ORR intermediates by the SACs and hence give rise to non-linear scaling relations while the similar mode of adsorption (top site) sustains it between OH* and OOH*. This observation immediately denies the possibility of considering O* binding energy as an activity descriptor which has been considered in many studies. To identify the chemical origin of breaking of scaling relations, we have carried out a projected crystal Hamilton population analysis (pCOHP) for all the intermediates, which has recently verified to be an important activity descriptor for computational screening studies (Figure 5).[39] The bonding population is represented in the right and antibonding population represented in left respectively.

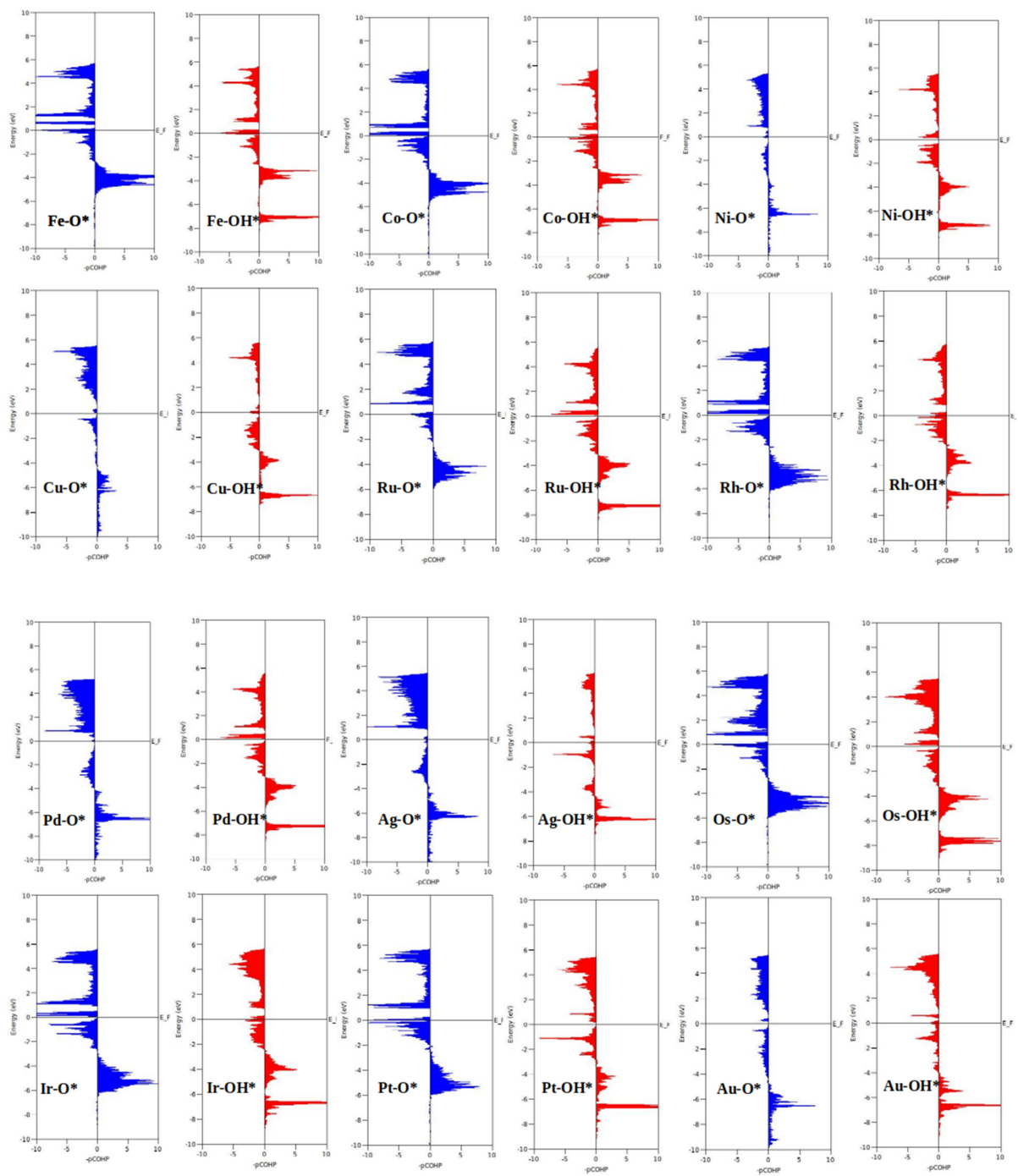

**Figure 5.** Projected crystal Hamilton population analysis of O* and OH* adsorbed on TM-Ph SACS.

The depicted O* pCOHP involves average of TM-O and P-O bonding for Ni, Cu, Pd, Ag, and Au based SACs as the O* occupies a bridging position in these systems. The observed discrepancies in the adsorption free energy and hence the scaling relation can be ascribed to the relatively higher variation in the spread of bonding and antibonding population for TM-O* binding as compared to TM-OH* binding. Another noteworthy aspect arising from the pCOHP analysis is that all the TM-Ph catalysts possess a slightly higher population for the TM-O* bonding energy levels than the antibonding energy levels at the valence band indicating an optimum bonding scenario which reduces the possibility of strong O* binding caused catalytic site poisoning, a major drawback of conventional transition metal based catalysts. Furthermore, the integrated COHP (ICOHP) up to the Fermi level plotted against the Mulliken charge transfer [40] reveals a higher order linearity between charge transferred to OH* compared to O* (Figure S3). While O* binding is associated with a larger charge span of 0.4-1.2|e| which is still far from the valence satisfying criteria of $2e^-$, OH* binding features a lower charge fluctuation (0.85-1.06|e|) which is consistent with the valency requirement of $1e^-$. This difference in the strengths of covalent character has been previously observed for OH* and OOH* for $MN_4$ complexes by Calle-Vallejo et al.[41] causing breaking of scaling relationships in vacuum and retained under implicit solvent consideration. Since our study involves an implicit solvation consideration, the differences in the covalence of ORR intermediates emerges as a parameter of interest for activity enhancement by tuning the metal center. The free energy diagrams for ORR (Figure 6) constructed at the equilibrium potential of 1.23 V advocates that there is a non-uniform trend in the energetics of ORR over different TM-Ph SACs.

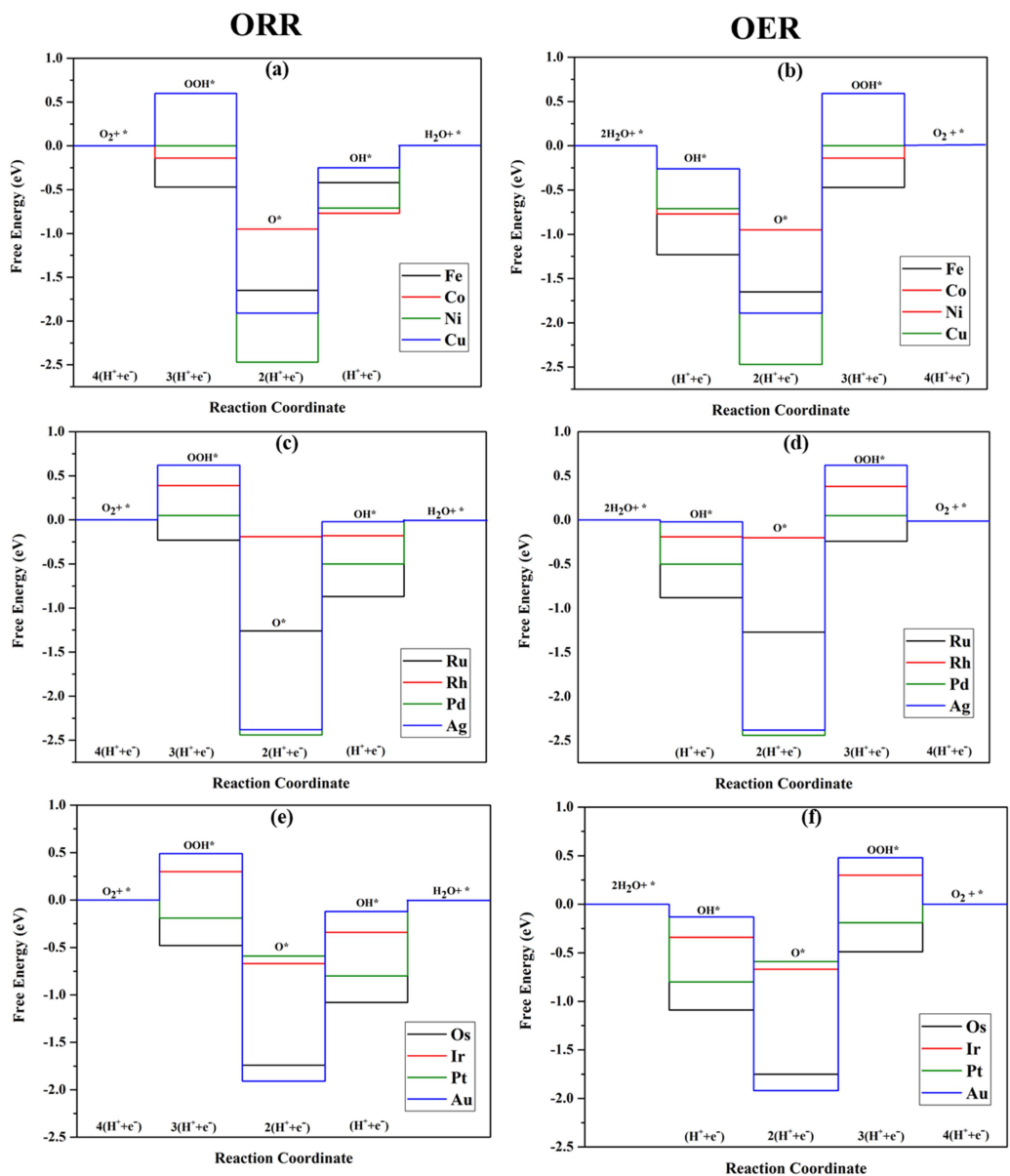

**Figure 6.** Free energy diagrams of ORR and OER along the TM-Ph SAC series constructed at 1.23V

The 3d series elements Fe, Ni and Cu along with end-lying Pd, Ag and Au of 4d and 5d series features OH* formation as the rate determining step. On the contrary, Co, Ru, Os, Ir and Pt based TM-Phs possess H$_2$O formation as the rate limiting step and Rh-Ph emerges as the only system with OOH* formation as rate limiting step. Therefore it is the stability of OH* on the catalyst surface which can be considered a crucial parameter in determining the catalytic trends among the SACs.

From the free energy screening, Ir-Ph turns out to be the best ORR catalyst with an overpotential of 0.34 V calculated using computational hydrogen electrode model (CHE) proposed by Norskov et al.,[42] which is followed by Rh and Co with overpotentials of 0.39 and 0.77 V respectively. It is interesting that Ir-Ph manifests an almost constant (although non-zero) ΔG (0.30, 0.37, 0.33 and 0.34 eV) for all the 4 protonation steps of ORR which is an adequate feature of an ideal catalyst. The lowest overpotential value of 0.34 eV reinforces the scaling relation driven activity trend by its excellent resemblance with the minimum thermodynamic overpotential of 0.33 V derived by the linear scaling between OH and OOH [(1.23 - (-4.92+3.12)/2)V =0.33 V]. Compared to ORR, OER activity follows a different trend across the SACs with a uniform rate determining step of OOH* formation. The highest activity has been observed for Pt-Ph for which a free energy change of 0.40 eV is required for the OOH* formation from O* which is followed by Rh-Ph with a corresponding value of 0.58 eV. The highest activity of Pt-Ph can be attributed to the $\Delta G_{O*}$ value of 1.87 eV lying closest to the mean value of $\Delta G_{OH*}$ and $\Delta G_{OOH*}$ which is an essential criteria for an efficient catalyst for which either O* to OOH* or OH* to O* being the rate determining step.[43] The dominance of single Pt atom compared to Ir or Ru for which the oxide counterparts are well-known OER catalysts in the activity regime is worthy to be noticed.

For both ORR and OER, high fluxionality is observed between the free energy of elementary reaction steps across the SACs with a notable uniformity in the end-lying metal center in the TM series (Cu, Ag, Au) exhibiting lowest activity. This can be ascribed to their high d electrons causing a strong binding towards the intermediates introducing energetically uphill trend in corresponding reactions. Activity contour plots of TM-Ph SACs for ORR and OER activity are represented in Figure 7(a-b) with $\Delta G_{OH*}$ and $\Delta G_{O*}-\Delta G_{OH*}$ as the reference parameters as the identified scaling between OOH* and OH* reduces the dependency of ORR/OER activity on four ΔG values into two. From the volcano plots, it is evident that the Ir-Ph, Rh-Ph, Pt-Ph and Os-Ph SACs occupy the highest activity regions and hence can be considered to be potential catalysts. It is also noteworthy that the lesser stable SACs Ag-Ph and Pd-Ph are also featured with a lesser electrocatalytic activity as identified from the activity volcano plot.

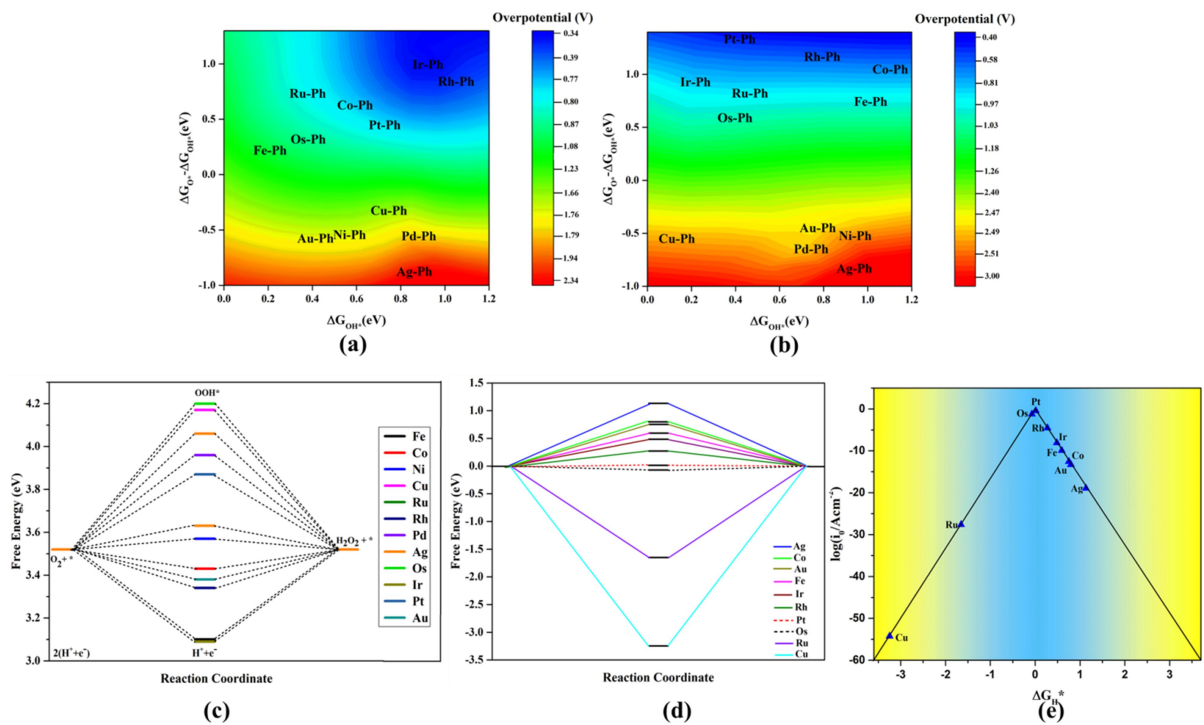

**Figure 7.** (a) Two dimensional ORR activity volcano plot, (b) two dimensional OER activity volcano plot, (c) free energy analysis of 2e⁻ reduction by TM-Ph SACs at 0.7 V, (d) HER free energy diagram of TM-Ph SACs, and (e) current density volcano plot of HER plotted against $\Delta G_{H*}$

To elucidate the thickness dependence on the electrocatalytic activity of TM-Ph based SACs, we have calculated the activity of Ir-Ph and Pt-Ph for ORR and HER, respectively upto a thickness (Table S3) of 3 monolayers (3MLs). Both the systems have shown a decrease in the activity with higher overpotentials for 2ML and 3ML thicknesses than the 1ML counterparts suggesting an inverse relation between ORR activity vs thickness. However, HER activity undergoes a comparatively lesser (by 1V) and irregular variation with thickness.

A competitive pathway for the 4e⁻ ORR with $H_2O$ formation is the 2e⁻ reduction resulting in $H_2O_2$ formation. Simultaneously, electrochemical synthesis of $H_2O_2$ is an important area of industrial importance.[44-46] Owing to the similarity in intermediates, we have investigated the 2e⁻ oxygen reduction activity of TM-Ph SACS by a free energy analysis with OOH* formation and $H_2O_2$ formation as the two elementary steps. Figure 7(c) represents the free energy diagram at the equilibrium potential of 0.7 V of 2e⁻ reduction. The ORR/OER active TM-Ph SACs have not been observed with prominent activity towards 2e⁻ reduction which diminishes the plausibility for competitive 2e⁻ vs 4e⁻ reduction pathways. Instead, 3d series based TM-Ph SACs such as Ni-Ph and Co-Ph emerge as active catalysts with a curtailed energy requisite of 0.05 and 0.09 eV leading to limiting potentials of 0.65 and 0.61 V, respectively. This deciphers an activity equivalence of the state of the art 2e⁻ reduction catalyst $PtHg_4$.[47] There is variation in the activity determining reaction from OOH* formation to OOH* protonation along the TM-Ph series occurring due to weak and strong binding of OOH* on the metal center, respectively.

Apart from oxygen reduction/evolution reactions, we have further investigated the hydrogen evolution activity of TM-Ph SACs as the recently developed many SACs have offered prevalent catalytic activity towards HER. Since Ni-Ph and Pd-Ph have been identified to be structurally unstable during H* adsorption, we have avoided them for HER activity study. Since $\Delta G_{H*}$ is the only activity descriptor involved, we have considered a Heyrovsky mechanism for our study subjecting to the fact that the protons are being adsorbed into the catalytic center from solution and later desorbed as $H_2$*.[48] From the HER free energy diagram given in Figure 7(d), it is obvious that majority of the SACs possess H* formation as the rate determining step whereas combination of adsorbed H* with protons to form $H_2$ is RDS for Cu and Ru. Here, Pt-Ph shows out to be the best SAC particularly almost achieving the ideal efficiency with $\Delta G_{H*}$ = 0.01 eV. This is followed by Os-Ph with an overpotential of 0.07 V. A volcano plot of exchange current density, a parameter which can be measured experimentally versus Gibbs free energy of adsorption represented in Figure 7e, where the current density is calculated by the following method proposed by Norskov et al.[49]

$$i_0 = \left\{\frac{-ek_0}{1+\exp\left[\frac{-G(H^*)}{k_bT}\right]}\right\} \text{ for } G(H^*) \leq 0 \text{ and } i_0 = \left\{\frac{-ek_0}{1+\exp\left[\frac{G(H^*)}{k_bT}\right]}\right\} \text{ for } G(H^*) \geq 0 \qquad (7)$$

where GH* is the free energy of H* adsorption, $k_b$ is the Boltzmann constant, T is the temperature and $k_0$ is the rate constant which has been taken to be 1 because of unavailability experimental results of TM-Ph SACs HER activity. Pt-Ph occupies at the topmost position of activity volcano suggesting the highest HER activity. Hence, the activity screening has conferred that phosphorene based SACs with transition metals belonging to group 9 and 10 emerge as highest active catalyst candidates (except Pd) considering all the electrocatalytic reactions. Also from our study, Pt-Ph emerges as a bifunctional catalyst with simultaneously offering the lowest Pt loading synergistic with excellent activity for OER and HER and can be expected to be a pivotal material for catalytic interest.

Phosphorene is well-known to be having stability issues at ambient conditions caused primarily by interaction with light, atmospheric oxygen and water/moisture. Being identified many of the TM-Ph SACs with exceptional electrocatalytic activity, we attempt to investigate the ambient condition stability of them via scrutinizing the chemical degradation of pristine phosphorene grounded from first principles calculations. The ambient condition stability investigations are carried out for the highest active catalysts Ir-Ph, Rh-Ph, Pt-Ph, Os-Ph for all the electrocatalytic reactions considered with pristine phosphorene as frame of reference.

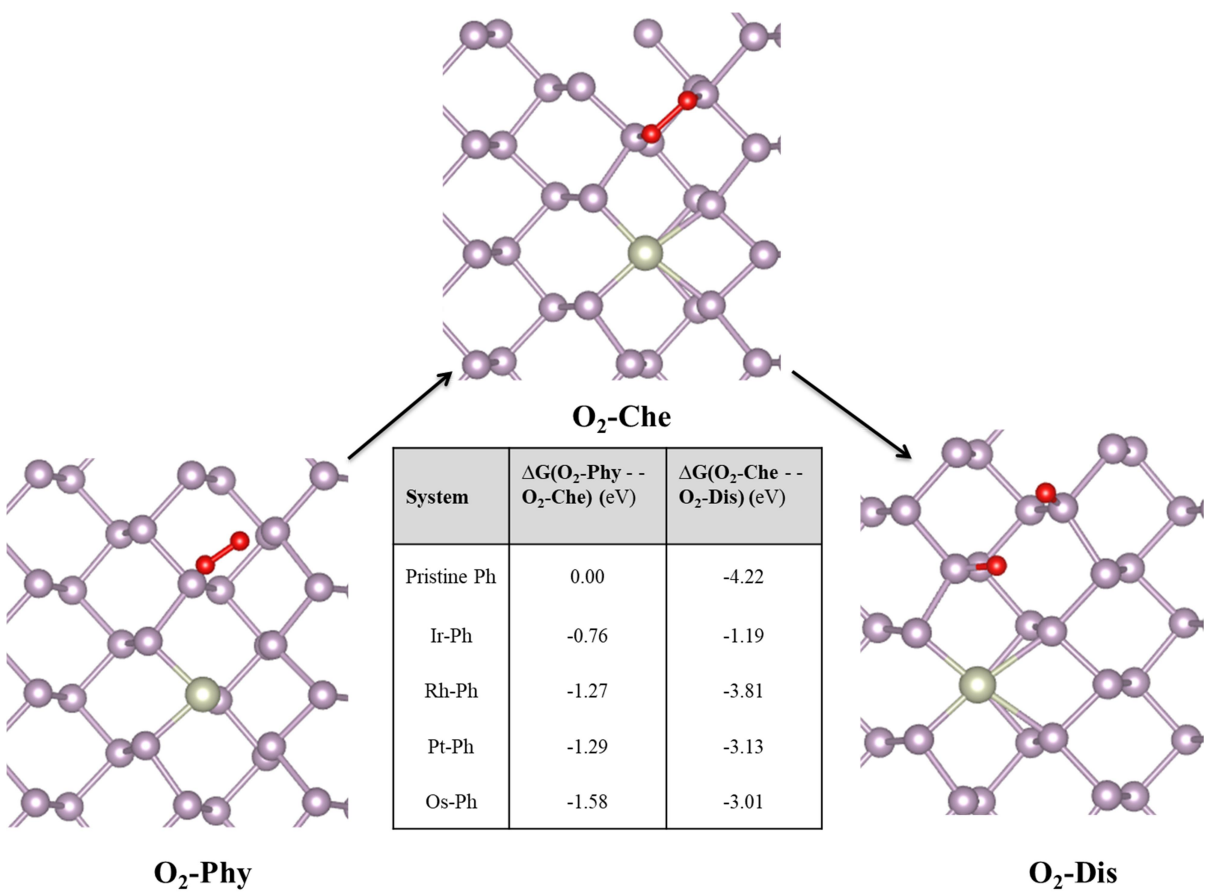

**Figure 8**. Free energy analysis of $O_2$ interaction on TM-Ph SACs. Pristine phosphorene is considered as the reference system. $O_2$-Phy, $O_2$-Che, $O_2$-Dis represents physisorbed, chemisorbed and dissociated oxygen, respectively.

Mechanism of $O_2$ induced degradation has been studied in detail by Ziletti et al. which is mediated by the dangling oxygen atoms bound to P atoms breaking the structural integrity of phosphorene monolayer.[50] We have extended the reported three different modes of $O_2$ interaction on phosphorene to the selected Ir-Ph, Pt-Ph, Rh-Ph and Os-Ph SACs which are i) $O_2$ physisorbed ($O_2$-Phy), ii) $O_2$ chemisorbed in bridge fashion ($O_2$-Che) and iii) $O_2$ dissociated into resulting in the formation of two P-O dangling bonds ($O_2$-Dis) on the non-TM region of TM-Ph systems. The structural details of three different adsorption modes are given on Table S4. Deviations in the magnetic moments from $2\mu B$ as observed for pristine phosphorene in some of the TM-Ph SACs suggest different interaction in the physisorption mode. The average binding energy of dangling O atoms for all the TM-Ph SAC except Pt-Ph are lower than that of pristine phosphorene implying a less stable 2 (P-O*) configurations for the respective SACs and hence a diminished affinity towards oxidation. Furthermore, we have investigated the free energy change associated with the transfer of $O_2$ interaction via the discussed three modes the results of which is represented in Figure 8. Although there is a thermodynamic plausibility for movement of $O_2$ from physisorbed to chemisorbed state, the exothermicity associated with $O_2$ dissociation is less for the SACs in comparison to the scenario of pristine phosphorene especially for the highest ORR active candidate Ir-Ph with the least exothermicity observed. This observation is supported by the low O* binding energy which can be attributed to the relatively larger occupation of antibonding energy levels under the valence band region for the least binding Ir-Ph SAC determined from the pCOHP analysis of P-O interaction as against pristine phosphorene (Figure S5). The previously observed transition metal states close to Fermi level revealed from the density of states analysis can be expected to reduce the pure p orbital character at least for the P atoms belonging to the immediate ridges and hence resulting in a lesser interaction with dissociated O* atoms. These results introduce a futuristic platform where the ambient condition stability can be tuned via more related chemical modification strategies by varying the composition, percentage of doping and synergistic methods with protective measures as well.

Another important degradation pathway for phosphorene is via the interaction with $H_2O$. Previous reports have shown that although phosphorene has fairly strong hydrophobicity towards $H_2O$, the oxygen bonded phosphorene is vulnerable to water catalyzed oxidation. Hereby, we analyze the interaction of water with O* bonded TM-Ph systems with pristine phosphorene considered as the reference system. The charge density difference analysis of the $H_2O$ interaction with O* bonded to the P atoms of TM-Ph systems is represented in Figure 9.

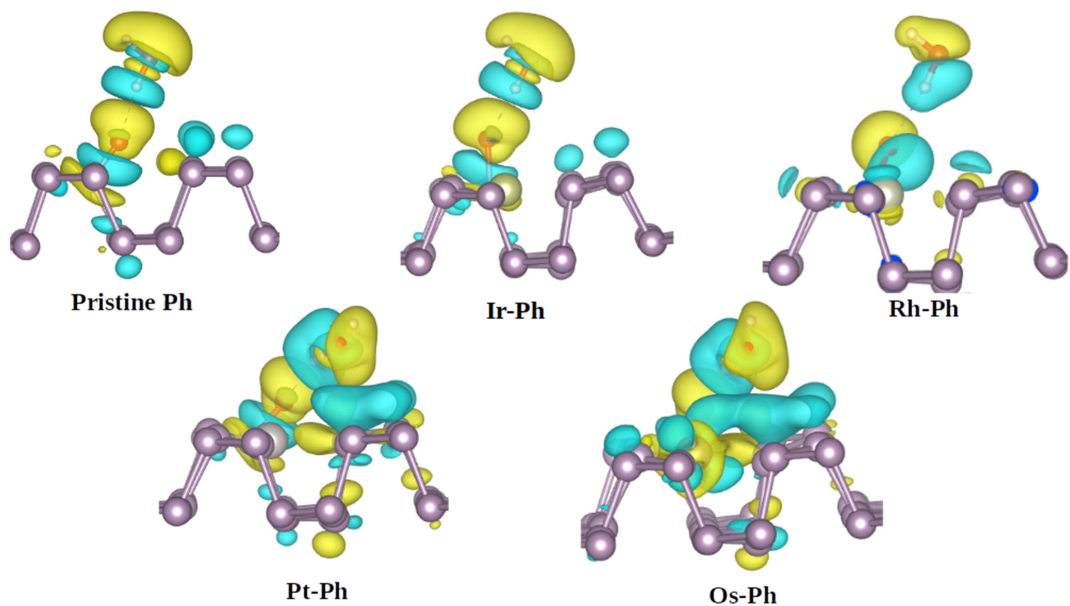

**Figure 9**. Charge density difference analysis of $H_2O$ with $O^*$ bonded TM-Ph SACs and pristine phosphorene. Yellow and Cyan colours represent positive and negative isosurfaces at 0.003eV.

It can be understood from the figure that while Ir-Ph shows a similar hydrogen bonding interaction to that of pristine phosphorene, Rh-Ph shows a less interaction with charges located majorly within water and the $O^*$ species only. Both Pt-Ph and Os-Ph interactions with $H_2O$ involve contribution from the TM atoms as well and also spreads over a larger number of P atoms across the phosphorene upper plane. This indicates that except Ir-Ph, all the TM-Ph systems exhibit a different extent of interaction with $H_2O$ with Rh-Ph exhibiting a diminished charge transfer which envisages the plausibility of controlling the $H_2O$ affinity of oxidized phosphorene via chemical modification.

In conclusion, we have systematically carried out catalytic activity of transition metal doped single atom catalysts (TM-Ph SACs) towards important catalytic reactions such as ORR, OER and HER. The stability analysis via binding energy, dissolution potential and electronic structure analysis confirms that all the TM-Ph SACs except Ag-Ph exhibiting excellent stability to withstand the electrochemical conditions. We have observed a breaking of scaling relationship between $O^*$ and $OH^*$ free energy of adsorption which is arising from the differences in the covalent characters attained by respective species across the SAC series. The free energy analysis suggest Ir-Ph:Rh-Ph, Pt-Ph:Os-Ph and Pt-Ph:Co-Ph as the most active catalyst systems towards ORR, OER and HER, respectively. 3d series based SACs Ni-Ph and Co-Ph are also found to be active catalysts for 2e$^-$ ORR. The ambient condition stability analysis focused on interactions of TM-Ph SACs with $O_2$ and $H_2O$ suggested that these interactions can be tuned by incorporating TM atoms into pristine phosphorene and alternative strategies based on this observation can be developed in future.

## ASSOCIATED CONTENT

**Supporting Information**. Structural details of transition metal doped phosphorene SACs, $O_2$ adsorption details, Projected Density of States (PDOS) of $O_2$ binding to Fe-Ph SAC, structural details of ORR intermediate binding, charge transfer and ICOHP analysis, ORR free energy diagrams of Ir-Ph SAC of thickness 2ML and 3ML, details of $O_2$ interaction with P atoms of TM-Ph SACs.

## AUTHOR INFORMATION

The authors declare no conflicts of interest

## ACKNOWLEDGEMENT

The authors acknowledge computing resources from the Swedish National Infrastructure for Computing (SNIC) at NSC and HPC2N. We thank IIT Indore for providing the lab/computational facilities and SPARC (SPARC/2018-2019/P116/S), DST SERB (CRG/2018/001131) projects for funding. ASN thank Ministry of Human Resources and Development, India for research fellowship.